\documentclass[apj,numberedappendix]{emulateapj}
\usepackage{graphics,epsf}
\usepackage{amsmath}                
\usepackage{amsfonts}               
\usepackage{amssymb}                
\usepackage{epsfig}                 
\usepackage{graphicx}
\usepackage{float} 
\usepackage{color}
\usepackage[para,online,flushleft]{threeparttable}

\newcommand{\km}{{~\rm km}}
\newcommand{\s}{{~\rm s}}

\newcommand{\K}{{~\rm K}}

\newcommand{\Myr}{{~\rm Myr}}

\newcommand{\kpc}{{~\rm kpc}}

\begin{document}

\title{The requirement for mixing-heating to utilize bubble cosmic rays to heat the intracluster medium}
\author{Noam Soker\altaffilmark{1,2}}

\altaffiltext{1}{Department of Physics, Technion -- Israel Institute of Technology, Haifa 32000, Israel;  soker@physics.technion.ac.il}
\altaffiltext{2}{Guangdong Technion Israel Institute of Technology, Shantou 515069, Guangdong Province, China}


\begin{abstract}
I conduct simple analytical estimates and conclude that mixing by vortices is a more efficient process to transfer the cosmic ray energy of jet-inflated bubbles to the intracluster medium (ICM) than streaming of cosmic rays along magnetic field lines.  Jets and the bubbles they inflate transfer heat to the ambient gas in cooling flows in cluster of galaxies and in galaxies. The internal energy of the jet-inflated bubbles is dominate by very hot thermal gas and/or cosmic rays. Cosmic rays that stream along magnetic field lines that connect the bubbles with the ICM heat the ICM as their energy is dissipated there. I find that about half of the cosmic ray energy is dissipated in the bubbles themselves. I also find that the ICM volume that the cosmic ray streaming process heats is only about five times as large as the volume of the bubbles. The outcome of heating by streaming only is that the cosmic rays form a larger bubble filled with very hot thermal gas. Therefore, there is a requirement for a more efficient process to transfer the internal energy of the bubbles to the ICM. I suggest that this process is heating by mixing that operates very well for both cosmic rays and the very hot thermal gas inside the bubbles. This leaves mixing-heating to be the dominant heating process of cooling flows. 
\end{abstract}


\section{INTRODUCTION}
\label{sec:intro}

Although the old ``cooling-flow problem'' in clusters and groups of galaxies, and in galaxies has been solved, there is still no consensus on the heating mechanism of the intracluster medium (ICM; for a recent review see \citealt{Soker2016}). 
The ICM in many clusters of galaxies has short radiative cooling times. If no heating processes of the ICM  exist large amounts of ICM should cool to low temperatures (by ICM I refer also to the interstellar medium in galactic cooling flows). 
The cooling-flow problem refers to the contradiction between the large mass cooling rates of the ICM that the no-heating assumption implies and the observations of much lower mass cooling rates. The solution is simply to relax the no-heating assumption and to consider the heating of the ICM in all cooling flows. The vast majority of heating models involve jets that the active galactic nucleus (AGN) at the center of the cooling flow launches. The jets and the hot low-density bubbles they inflate heat the ICM via a negative feedback mechanism that prevents a run-away heating or cooling (e.g., \citealt{Fabian2012, Farage2012, McNamaraNulsen2012, Gasparietal2013, Pfrommer2013, Baraietal2016, Soker2016, Birzanetal2017, Iqbaletal2017}). 

The two parts that compose the negative feedback cycle are the feeding of the AGN by the ICM and the heating of the ICM by the AGN. There are tens of observational and theoretical studies in recent years that support models in which the ICM feed the AGN with cold clumps (e.g., limiting the list to the last 3 years, \citealt{ChoudhurySharma2016, Hameretal2016, Loubseretal2016, McNamaraetal2016, Russelletal2016, Tremblayetal2016, YangReynolds2016b, Davidetal2017, Donahueetal2017, Gasparietal2017, Hoganetal2017, Meeceetal2017, Prasadetal2017, Russelletal2017a, Voitetal2017, Babyketal2018, Gasparietal2018, Jietal2018, Prasadetal2018, Pulidoetal2018, Vantyghemetal2018, Voit2018, YangLetal2018}). 
These models can be grouped under the \emph{cold feedback mechanism} that \cite{PizzolatoSoker2005} developed to replace the then popular Bondi accretion. In the cold feedback mechanism gas flows inward to form cold clouds and to feed the central AGN. Cold clumps are observed now down to the AGN, e.g., in Perseus \citep{FujitaNagai2017}. This implies that a cooling flow does take place, like the extreme cooling flow in the Phoenix cluster (e.g., \citealt{Pintoetal2018}), although in most cases the cooling flow is a moderate one. For that, in this study I use the term cooling flow (rather than other names that were invented later and caused some confusions; \citealt{Soker2010}). 

On the other hand, there is no agreement on the processes that contribute the most to the heating of the ICM by the jets that the AGN launches and the bubbles that they inflate. I find it useful to distinguish between heating processes where the jets and jet-inflated bubbles do work on the ICM and between heating by energy transport. In the first class of processes the jets and the jet-inflated bubbles do work on the ICM by exciting sound waves (e.g., \citealt{Fabianetal2006, Fabianetal2017, TangChurazov2018}), by driving shocks (e.g., \citealt{Formanetal2007, Randalletal2015, Guoetal2018}), by powering turbulence (e.g.,    \citealt{DeYoung2010, Gasparietal2014, Zhuravlevaetal2014, Zhuravlevaetal2017}), and/or by uplifting gas from inner regions (e.g., \citealt{GendronMarsolaisetal2017}). 
However, there are studies that show that these processes do not heat the ICM efficiently enough, despite the fact that these processes themselves take place to some degree. Although turbulence is observed in the ICM (e.g., \citealt{Zhuravlevaetal2014, Zhuravlevaetal2015, AndersonSunyaev2016, Arevalo2016, Hitomi2016, Hofmannetal2016,Hitomi2017}), there are studies that question turbulent heating (e.g., \citealt{Falcetaetal2010, Reynoldsetal2015, Hitomi2016, HillelSoker2017a, Bambicetal2018, MohapatraSharma2018}). Sound waves also are unable to supply the entire heating (e.g. \citealt{FujitaSuzuki2005}). Heating by shocks has some problems as well (e.g., \citealt{Sokeretal2016}). Observations show that bubbles can uplift cooler gas (e.g,. \citealt{Russelletal2017b, Suetal2017, GendronMarsolaisetal2017}), and some studies simulate the uplifting process (e.g., \citealt{Guoetal2018, Churazovetal2001}), but \cite{HillelSoker2018} claim that this cannot be the main heating process of the ICM. 

In the second class of heating processes the energy from the hot jet-inflated bubbles is carried into the ICM. One possibility, that is the main subject of the present study, is that cosmic rays that are accelerated within the jet-inflated bubbles, stream into the ICM and heat it (e.g. \citealt{Fujitaetal2013, FujitaOhira2013, Pfrommer2013}).
The second possibility is heating by mixing that works as follows. As the jets propagate through the ICM and inflate bubbles they form many vortices inside the bubble and on the boundary between the ICM and the bubbles. These vortices mix hot bubble gas into the ICM (e.g., \citealt{BruggenKaiser2002,  Bruggenetal2009, GilkisSoker2012, HillelSoker2014, YangReynolds2016b}).  

Finally, some studies suggest that two or more processes out of the different process can work together,  e.g., thermal conduction and cosmic rays (e.g., \citealt{GuoOh2008}), mixing of cosmic rays that were accelerated inside jet-inflated bubbles with the the ICM, \citep{Pfrommer2013}, and heating by turbulence together with turbulent-mixing (e.g. \citealt{BanerjeeSharma2014}).
 
In a series of papers (e.g., \citealt{HillelSoker2017a, HillelSoker2018} for most recent papers) we argued that although the inflation of bubbles also excite sound waves, shocks, and ICM turbulence, heating by mixing is much more efficient than the heating mechanisms where the jets and bubbles do work on the ICM. The heating by mixing can work for bubbles that are filled with very hot thermal gas or with cosmic rays, or a combination of the two.
I consider these simulations to describe the inflation of bubbles by jets and the bubble evolution even when weak magnetic fields are presence. Although magnetic fields can change the details of the interaction of the bubbles with the ICM (e.g., \citealt{DursiPfrommer2008, Weinbergeretal2017}), I take the view that the main effect is that the bubbles amplify the ICM magnetic fields (e.g., \citealt{DursiPfrommer2008, Soker2017}). For example, magnetic fields can suppress the Kelvin-Helmholtz instability along the field lines (e.g., \citealt{DursiPfrommer2008, Weinbergeretal2017}), but not perpendicular to the field lines. Therefore, magnetic fields are expected to change the geometry of the instabilities and to some extend the geometry of vortices that mix the ICM and bubbles, but magnetic fields are not expected to change the existence and global roles of the vortices, that include mixing and the amplification of the magnetic fields. 

 We did not compare the heating by mixing with the heating process where cosmic rays from the bubbles stream, instead of being mixed by vortices, into the ICM and heat it. 
Observations show that in older bubbles the contribution from sources other than the cosmic ray electrons to the pressure tends to be larger than in younger bubbles (e.g., \citealt{DunnFabian2004, Birzanetal2008, Crostonetal2008}). According to the results of the present study this additional pressure component could result from thermal hot gas heated by the mixing-heating process. The mixing-heating proceeds and increases the amount of thermal gas on the expense of the cosmic ray energy as the bubbles age. 
    
 In light of some recent interest in cosmic ray heating by streaming (e.g., \citealt{Ruszkowskietal2017, Ehlertetal2018, JiangOh2018, Ruszkowskietal2018}), and the possibility that in many cases cosmic ray energy is the main energy content of bubbles (e,g., \cite{Abdullaetal2018} for MS~0735.6+7421), in this study I examine some of the properties of this process. In section \ref{sec:heating} I consider only cosmic ray streaming, without mixing by vortices, and study the partition of cosmic ray energy between heating the bubbles themselves and heating the ICM, and in section \ref{sec:ICMvolums} I estimate the volume of the ICM that can be heated by streaming cosmic rays. I discuss and summarize my results in section \ref{sec:summary} where I consider mixing of cosmic rays by vortices to be much more efficient than streaming. 
   
\section{Heating the Bubble and the ICM}
\label{sec:heating}
 
  The basic assumptions of the cosmic ray heating are as follows.
 The heating per units volumes is (e.g., \citealt{Ehlertetal2018})
\begin{equation}
\mathcal{H}_{\rm cr} = \vert v_A \cdot \overrightarrow {\nabla} P_{\rm cr} \vert, 
\label{eq:jet}
\end{equation}
 where $v_A=B/\sqrt {4 \pi \rho}$ is the Alfven speed, $\rho$ is the density, and $P_{\rm cr}$ is the cosmic ray pressure.
I consider only cosmic ray streaming heating as it dominates over Coulomb and hadronic heating (e.g., \citealt{Ruszkowskietal2017}).
 
 Consider an ideal situation where the bubble is filled with cosmic rays that are disconnected from the ICM by magnetic fields. Then in a short time magnetic field lines from the bubble reconnect to the magnetic field lines in the ICM, and the cosmic rays stream to the ICM and heat it. Namely, there is a magnetic flux tube of cross section $S$ that connects the bubble to the ICM.  

\begin{figure}
 \begin{center}
  \hskip -0.6 cm
 \includegraphics[trim= 1.9cm 0.9cm 1.9cm 0.8cm,clip=true,width=0.50\textwidth]{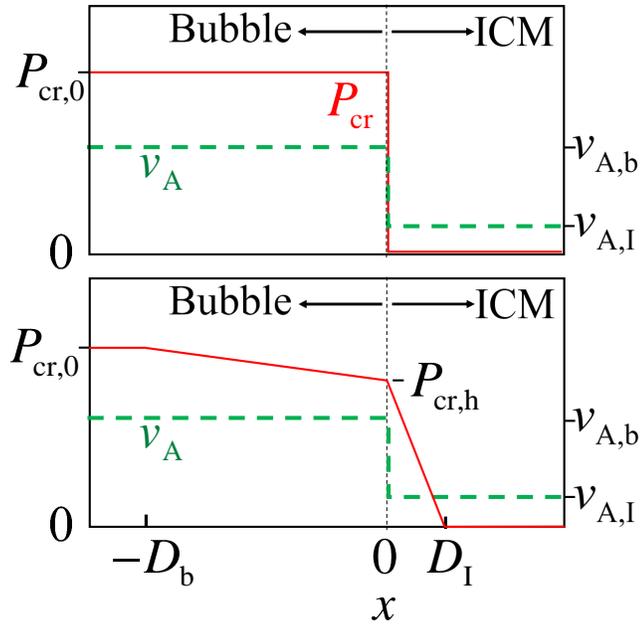}
\vskip -4.4 cm
\caption{A schematic drawing of the contact between the hot bubble, on the left of the dashed vertical line, and the ICM, as marked. The $x$ coordinate is along the magnetic flux tube that connects the bubble to the ICM. The upper panel represents the case at $t=0$ when reconnection of the magnetic field lines inside the bubble and in the ICM takes place. The lower panel represents the flow at some time later. The green-dashed and red-solid lines depict the Alfven speed (scale on the right) and the cosmic ray pressure (scale on the left), respectively.   }
\label{fig:schematic}
\end{center}
\end{figure}

Let as take the geometry as \cite{Ehlertetal2018} obtain in their simulations, where the cosmic rays leaks from the bottom of the bubbles in a pseudo-cylindrical stream. Namely, the cross section of the cosmic ray stream stays about constant.    
We can treat then the flow as a one-dimensional flow. The situation before reconnection is depicted schematically in the upper panel of Fig. \ref{fig:schematic}. I take the $x$ coordinate to be along the magnetic field lines, with $x=0$ at the boundary between the bubble and the ICM.  
 The initial cosmic ray pressure inside the bubble is $P_{\rm cr,0}$ while in the ICM it is zero. 
 The magnetic field in the bubble, $B_ {\rm b}$, is expected to be stronger than that in the ICM, $B_{\rm I}$, and the density inside the bubble much smaller than that in the ISM. As such, the Alfven speed inside the bubble is much larger than that in the ICM $v_{\rm A,b} \gg v_{\rm A,I}$.  
 
 The lower panel of Fig. \ref{fig:schematic} schematically presents the flow some time after reconnection. The value of the cosmic ray pressure in the boundary between the bubble and the ICM decreases to a value of $P_{\rm cr} (x=0) =P_{\rm cr, h}$, and it extends to distance $D_{\rm I}$ into the ICM, i.e., 
 $P_{\rm cr} (x=D_{\rm I}) =0$, while the reduction in cosmic ray pressure  extends to a distance of $D_{\rm b}$ into the bubble. 
 
I assume that the Alfven speed in the ICM is constant, and so is the Alfven speed in the bubble, but the later is larger. The cosmic ray heating rate of the ICM per unit area, where the units area is  perpendicular to the magnetic field lines, is
\begin{equation}
\dot E_{\rm I} = \int_0 ^{D_{\rm I}} 
\mathcal{H}_{\rm cr} dx  
\simeq  v_{\rm A,I} \int_0 ^{D_{\rm I}} \left|  \frac{d P_{\rm cr}}{dx} \right| dx 
\simeq v_{\rm A,I} P_{\rm cr, h} . 
\label{eq:HICM1}
\end{equation}
In the same manner, the heating rate per unit area inside the bubble is 
\begin{equation}
\dot E_{\rm b} = \int_{D_{\rm b}}^0 
\mathcal{H}_{\rm cr} dx  
\simeq v_{\rm A,b} \left( P_{\rm cr, 0} -P_{\rm cr, h} \right). 
\label{eq:Hb}
\end{equation}
 
 I now further assume that in a short time following the reconnection of the bubble and the ICM field lines, the value of the cosmic ray pressure at $x=0$ drops from $P_{\rm cr, 0}$ to a more or less constant value of $P_{\rm cr, h}$. The value of $P_{\rm cr, h}$ will continue to decrease after the information of the dissipation of cosmic ray has reached the other side of the bubble, 
$t_{\rm b} = 2 R_{\rm b}/_{\rm A,b} $, where  $R_{\rm b}$ is the radius of the bubble. For a time of $t \la t_{\rm b}$ I take the value of $P_{\rm cr, h}$ not to change much. The energy of the cosmic ray that has been dissipated is supplied by cosmic ray energy flux from the bubble $F_{\rm cr} \simeq (e_{\rm cr} + P_{\rm cr}) v_{\rm A,b}=  4 P_{\rm cr} v_{\rm A,b} $, where $e_{\rm cr} = 3  P_{\rm cr}$ is the  cosmic ray energy density inside the bubble. 
I calculate the cosmic ray energy that the bubble supplies over time $t$ by calculating the decrease of the cosmic ray energy and pressure inside the bubble per unit area.
For an approximate linear relation between the cosmic ray pressure inside the bubble and distance into the bubble, as I mark in Fig. \ref{fig:schematic}, over a time $t$ this flux has supplied a cosmic ray energy per unit area of 
\begin{equation}
E_{\rm cr, sup} \simeq 
\left( 4 P_{\rm cr, 0} - 4 P_{\rm cr, h} \right) 
\frac{ D_{\rm b} }{2}
\simeq 2 \left( P_{\rm cr, 0} -  P_{\rm cr, h} \right) v_{\rm A,b} t, 
\label{eq:Esupply}
\end{equation}
where I take $D_{\rm b} \simeq v_{\rm A,b} t$. 
The dissipated cosmic ray energy per unit area at the same time is given by 
 \begin{equation}
E_{\rm cr, dis} =  \left( \dot E_{\rm I}+\dot E_{\rm b} \right)t .
\label{eq:Edissipate}
\end{equation}
Equating the dissipated energy to the supplied energy $E_{\rm cr, dis}= E_{\rm cr, sup}$ and using equations (\ref{eq:HICM1}) and (\ref{eq:Hb}), gives the relation 
\begin{equation}
\left( P_{\rm cr, 0} -  P_{\rm cr, h} \right) v_{\rm A,b} 
\simeq  P_{\rm cr, h} v_{\rm A,I} .
\label{eq:EsupEdis}
\end{equation}
 
 Now we can use back equations (\ref{eq:HICM1}) and (\ref{eq:Hb}), and conclude from equation (\ref{eq:EsupEdis}) that the cosmic ray dissipation rate inside the bubble is about equal to that in the ICM 
\begin{equation}
\dot E_{\rm I} \simeq \dot E_{\rm b}. 
\label{eq:EIEb}
\end{equation}
 
It is important to note that both the Alfven velocity in the bubble and in the ICM have been canceled out from the final expression (\ref{eq:EIEb}). In particular, even if the Alfven velocity inside the bubble is relativistic equation (\ref{eq:EIEb}) holds. This is true also in the case that the information inside the bubble travels at a speed of $c/\sqrt{3}$, where $c$ is the speed of light, as argued for by \cite{ThomasPfrommer2018}.  
 
I examine some recent works on cosmic ray heating in light of equation (\ref{eq:EIEb}). 
Equation (\ref{eq:EIEb}) implies that one cannot ignore the cosmic ray energy that is transferred to heat the jet-inflated bubbles themselves. \cite{Ehlertetal2018}, for example, turn off the cosmic ray dissipation in regions where the matter from the bubble dominate. I find this unjustified.
\cite{JacobPfrommer2017a} and \cite{JacobPfrommer2017b} derive the cosmic ray heating from a steady-state solution to clusters. They do not include the inflation of bubbles self-consistently in their treatment, and therefore I claim that they substantially underestimate the dissipation of cosmic ray inside the bubbles and near the bubble surface. Namely, they substantially underestimate the dissipation of cosmic ray energy near the source of the cosmic rays. 
 
\cite{Ruszkowskietal2017} present simulations of feedback heating by cosmic rays. Unfortunately they do not present the evolution with time of the velocity flow of the ICM, nor the magnetic field in the regions where they inject the jets. Therefore, I cannot use their results to estimate the role of mixing or cosmic ray heating inside the hot bubbles. But I do note that they assume that a fraction of $f_{\rm cr} = 0.8$ of the energy of the jets ends as cosmic rays. Equations (\ref{eq:EIEb}) implies that in that case, at most a fraction of $0.5 f_{\rm cr}=0.4$ of the jets' energy will end up as cosmic ray heating of the ICM. But I argue in this paper that most of the cosmic ray energy is carried to the ICM by mixing, and not by streaming. 
    
I neglected above the process by which turbulence inside the bubble can smooth the cosmic ray pressure, and by that reducing the dissipation of cosmic ray energy inside the bubble. However, this does not seem to be a problem to the derivation of equation (\ref{eq:EIEb}). Firstly, I do think that there is a strong turbulence, including large vortices, inside the bubble, as the vortices play a significant role in the heating by mixing process, which I take to be the most significant heating process.  Secondly, turbulence entangles magnetic field lines that reconnect. This process prevents a smooth flow of cosmic rays into the ICM, increasing further the problem that the cosmic rays do not heat a large enough volume of the ICM. I turn to discuss this problem next. 

\section{The heated ICM volume}
\label{sec:ICMvolums}

The front of the cosmic rays advances at a speed of $\simeq  v_{\rm A,I}$ into the ICM. The speed might be higher than this by at most few tens of percents (e.g., \citealt{Ruszkowskietal2017}). At the same time the information that cosmic ray energy is lost from the bubble travels at a speed of $\simeq  v_{\rm A,b}$ into the bubble. When this information covers the entire volume of the bubble $V_{\rm b}$, at time $\approx t_{\rm b}$, the volume of the ICM that was heated is 
\begin{equation}
V_{\rm I} (t_{\rm b}) \approx  V_{\rm B} \frac {v_{\rm A,I}}{v_{\rm A,b}}.
\label{eq:VolumeB}
\end{equation}
I calculate the energy of the cosmic ray in the bubble that supplies the dissipated cosmic ray energy from equation (\ref{eq:Esupply}). I do this by substituting $t=t_{\rm b}$ and taking $ V_{\rm b}={v_{\rm A,b}} t_{\rm b} S$ to be the volume of the bubble, where $S$ is the cross section of the magnetic flux tube that connects the bubble to the ICM.  
This energy that the cosmic rays supply is a fraction $\eta$ of the entire cosmic ray energy that the bubble can supply, where $\eta$ is given by  
\begin{equation}
\eta =\frac {S E_{\rm cr, sup} (t_{\rm b}) }{ 4 P_{\rm cr, 0} V_{\rm b}}  \simeq 
\frac{ P_{\rm cr, 0} -  P_{\rm cr, h} } {2 P_{\rm cr, 0}}.  
\label{eq:eta}
\end{equation}

I crudely take the time to extract most of the available cosmic energy from the bubble to be $\approx t_{\rm b} /\eta$. The ICM volume that is covered with cosmic ray heating by streaming, until the entire bubble cosmic ray energy is consumed, is   
\begin{equation}
V_{\rm I, heat} \approx \frac{t_{\rm b}/\eta}{t_{\rm b}} V_{\rm I}(t_{\rm b}) 
\approx 
 V_{\rm B} \frac {v_{\rm A,I}}{v_{\rm A,b}}
\frac{2 P_{\rm cr, 0}}{P_{\rm cr, 0} -  P_{\rm cr, h} } 
\approx 4 V_{\rm B},
\label{eq:VolumeI}
\end{equation}
where in the second equality I substituted from equations (\ref{eq:VolumeB}) and (\ref{eq:eta}), and in the third equality I substitute equation (\ref{eq:EsupEdis})  with the approximation that the average value of $P_{\rm cr, h}$ is  
$\overline P_{\rm cr, h} = 0.5 P_{\rm cr, 0}$. 

The approximate equation (\ref{eq:VolumeI}) together with equation (\ref{eq:EIEb}) imply that by streaming alone the cosmic ray energy is channeled to thermal energy that heats a region of approximately only five times the initial volume of the bubbles. From then on, we have larger hot bubbles that still need to heat most of the ICM. There must be another much more efficient process to transfer the energy from the hot bubbles to the ICM. 

I further clarify two issues related to the simple model that I have been using here. I studied the case of a bubble that is no longer energized by jets. A more thorough study is required for the less ideal case where the jets continuously pump energy into the bubble. I do not expect the conclusions to change much.    
 
I also assumed above that the cosmic rays stream along magnetic filed lines. Cosmic rays can also diffuse between field lines. The diffusion between field lines is a slow process and might make the cosmic ray distribution less steep, resulting in a slower energy dissipation in the ICM. However, for the cosmic ray streaming to be significant in the heating process it must transfer the cosmic ray energy before the bubbles move outward. 
Let me consider an example. 

In the simulation of \cite {Ehlertetal2018} at $t=60 \Myr$ the lower part of the bubble is at $r \simeq 60 \kpc$ and the upper boundary of the bubble is at $r \simeq 90 \kpc$. This implies that the bubble leaves the center of the cluster, $r \la 30 \kpc$, at a speed of $v_{r,{\rm bubble}} \la 1000 \km \s^{-1}$. The cosmic rays stream at about the Alfven speed. Since the magnetic pressure in the ICM is typically less than $10 \%$ of the thermal pressure, the Alfven speed is at most 30 per cent the sound speed. Namely,  $v_{\rm A} \la 0.3 c_s \approx 200 \km \s^{-1} \ll v_{r,{\rm bubble}}$. If the cosmic rays diffusion time is slower than what I assume in the present study, then the cosmic rays will not have time to diffuse out from the bubbles and heat the center of the cluster before the bubbles move out. It is in the center of the cluster where the heating is most needed. 

In that regards I notice that \cite{Sharmaetal2009a} argued that the cosmic-ray diffusion in the ICM takes place on a longer time scale than the buoyancy time. They (see also \citealt{Sharmaetal2009b}) also argued that turbulent mixing is more efficient than diffusion in transporting cosmic ray energy in the ICM. My results on the unitizing of the cosmic ray energy inside bubbles is compatible with their findings for the transport of cosmic ray energy in the ICM.  

\section{DISCUSSION AND SUMMARY}
\label{sec:summary}

By means of simple analytical estimates I examined the efficiency by which streaming of cosmic rays along magnetic field lines can transfer  energy from the jet-inflated bubbles to the ICM.  In section \ref{sec:heating} I found that about half of the cosmic ray energy is dissipated in the bubbles themselves. Even before the dissipation of cosmic ray energy starts, some of the kinetic energy of the jets has been converted to thermal energy of very hot gas inside the bubbles. Over all, more than half of the kinetic energy of the jets ends up as very hot thermal gas inside the bubbles.   
In section \ref{sec:ICMvolums} I found that by streaming alone the cosmic rays heat a total ICM volume of about five times the volume of the bubbles, and hence this mechanism does not cover a large enough volume of the ICM for an efficient heating to work. 
  
 Taking these two conclusions together, I argue that the effect of cosmic ray heating by streaming alone turns a bubble filled with cosmic rays to a larger bubble filled with very hot thermal gas. Therefore, although heating by cosmic ray streaming does take place, I think there is a need for an additional and a more efficient process to carry the energy from the bubbles to the ICM. 
  
I take this more efficient process to be heating by mixing (for references see section \ref{sec:intro}). The propagation of jets through the ICM and the inflation of the bubbles form many vortices in the ICM and in the bubbles, and these mix the content of the bubbles, i.e., the very hot thermal gas, the cosmic rays, and the magnetic fields, with the ICM. 
 When the bubble content and the ICM are well mixed we expect that magnetic fields from the bubble content and from the ICM reconnect on small scales. Then streaming of cosmic rays on small scale can make the final cosmic ray energy transfer, as much as heat conduction can locally transfer energy from the very hot thermal gas, which is now well mixed with the ICM,  to the ICM.  
 
The heating by mixing process that I refer to results from the vortices that the inflation of bubbles form (e.g., \citealt{GilkisSoker2012, HillelSoker2016}), and it is indiscriminate to the content of the bubble, and hence works for cosmic rays as well as for the thermal gas. Indeed, \cite{Pfrommer2013} already mentioned the mixing of cosmic rays with the ICM and referred to it as an important process. Here I argue that it is more significant than heating by cosmic ray streaming, and hence I strengthen our earlier claim that heating by mixing is the main process by which jets-inflated bubbles heat the ICM (see section \ref{sec:intro}).   

I emphasize that the non-relativistic thermal gas inside bubbles must be very hot, $T \ga 10^9 \K$. There are strong observational limits on the mass inside the bubbles (e.g., \citealt{SandersFabian2007, Abdullaetal2018}), and pressure balance between the bubbles and the ICM requires therefore the gas inside the bubbles to be very hot. Both mixing of the ICM with cosmic rays and inflation of the bubbles with jets at velocities of $\ga 10^4 \km \s^{-1}$ can account for this very hot gas. 
Another observational constrain comes from observations of cold rims around some bubbles (e.g., \citealt{Blantonetal2003}). Our 3D hydrodynamical simulations show that a cold dense gas around rising bubbles exists alongside the operation of the mixing-heating process \citep{HillelSoker2018}. More studies are required to better compare heating-mixing with these and other observations, e.g., radio observations. These studies require new sets of 3D magnetohydrodynamical simulations.   

The simulation of the feedback process in cooling flows is a very complicated task, e.g., as \cite{Martizzietal2018} show in a very recent study. There are many physical and numerical issues to consider. One of the key ingredients that simulations must include is the launching of kinetic jets. \cite{SternbergSoker2008} show that injecting energy off-center instead of launching jets might miss key processes, such as the formation of many vortices. It might be that by inserting their jets in a sphere off-center  \cite{Weinbergeretal2017}  and \cite{Ehlertetal2018} underestimated the role of heating by mixing \citep{HillelSoker2017b}.

The heating by mixing also has some implications to the mass feeding part of the feedback cycle. \cite{PizzolatoSoker2005} showed that non-linear perturbations in the ICM are required to form the cooling clumps that feed the AGN, as was also confirmed in following studies (e.g., \citealt{Gasparietal2018}). \cite{PizzolatoSoker2005} also suggested that the AGN activity form these non-linear perturbations. This is most likely a result of the vortices and the mixing process. Therefore, in an indirect way, the results of the present study that further strengthen the mixing-heating mechanism, add also to the cold feedback mechanism. 

I thank Karen Yang, Yutaka Fujita, Prateek Sharma, Christoph Pfrommer and an anonymous referee for useful comments.  
This research was supported by the Israel Science Foundation, and by the E. and J. Bishop Research Fund at the Technion.

\label{lastpage}
\end{document}